\documentstyle[12pt]{article}
\begin{document}
\begin{titlepage}
\begin{flushright}
Lecce University LE/ASTRO-1/97\\
Pavia University FNT/T-97/4\\
Zurich University ZU-TH 1/97\\
\end{flushright}
\vfill
\begin{center}
{\large\bf HALO DARK CLUSTERS OF BROWN DWARFS AND MOLECULAR CLOUDS}
\vskip 0.5cm
F.~De Paolis$^{1}$,
G.~Ingrosso$^{2,*}$ ,
Ph.~Jetzer$^{1}$
and M.~Roncadelli$^{3,\dagger}$
\vskip 0.5cm
$^1$
Paul Scherrer Institute, Laboratory for Astrophysics, CH-5232 Villigen PSI, and
Institute of Theoretical Physics, University of Zurich, Winterthurerstrasse
190, CH-8057 Zurich, Switzerland\\
$^2$
Dipartimento di Fisica, Universit\`a di Lecce, Via Arnesano, CP 193, I-73100
Lecce, and INFN, Sezione di Lecce, Via Arnesano, CP 193, I-73100 Lecce, Italy\\
$^3$
INFN, Sezione di Pavia, Via Bassi 6, I-27100, Pavia

\vfill
\end{center} 
 
\begin{abstract}
\noindent
The observations of microlensing events in the Large Magellanic Cloud
suggest that a sizable fraction ($\sim$ 50\%)
of the galactic halo is in the form of
MACHOs (Massive Astrophysical Compact Halo Objects) with an 
average mass
$\sim 0.27 M_{\odot}$, assuming a standard spherical
halo model. We describe 
a scenario in which dark clusters of MACHOs and 
cold molecular clouds (mainly of $H_2$) naturally form in the halo at 
galactocentric distances larger than 10-20 kpc. 
\end{abstract}
\vfill
\begin{flushleft}
$^*$ Partially supported by Agenzia Spaziale Italiana.\\
$^{\dagger}$ Partially supported by Dipartimento di Fisica Nucleare
e Teorica, Universit\`a di Pavia, Pavia, Italy.\\
Talk presented by Ph. Jetzer at the 18th Texas Symposium
on Relativistic Astrophysics
(Chicago, December 1996). To appear in the proceedings.
\end{flushleft}
\end{titlepage}
\newpage 
\baselineskip=21pt
\section{Introduction}
A central problem in astrophysics concerns the nature of the
dark matter in galactic halos, whose presence is implied by the flat 
rotation curves in spiral galaxies. As first proposed by 
Paczy\'nski \cite{pa}, 
gravitational microlensing can provide a decisive answer to 
that question \cite{kn:Derujula1},
and since 1993 this dream has started to become a reality 
with the detection of several microlensing events towards the Large 
Magellanic Cloud \cite{al,au}. 
Today, although the evidence for 
MACHOs is firm, the 
implications of this discovery crucially depend on the assumed galactic 
model. It has become customary to take the standard spherical halo model as 
a baseline for comparison. Within this model, the mass moment method yields 
an average MACHO mass \cite{je} of $0.27~M_{\odot}$. 
Unfortunately, because of the presently available limited statistics 
different data-analysis procedures
lead to results which are only marginally consistent. 
For instance, the average mass reported by the MACHO team is
$0.5^{+0.3}_{-0.2}~M_{\odot}$. Apart from the low-statistics problem  --
which will automatically disappear from future larger data samples -- we 
feel that the real question is whether the 
standard spherical halo model correctly describes our galaxy
\cite{kn:Ingrosso}.
Besides the observational evidence that spiral galaxies generally have 
flattened halos, recent determinations of the disk scale length, the 
magnitude and slope of the rotation at the solar position indicate that
our galaxy is best described by the maximal disk model.
This conclusion is further strengthened 
by the microlensing results towards the galactic centre, which 
imply that the bulge is more massive than previously thought.
So, the expected average MACHO mass should be smaller than within the
standard halo model. Indeed, the value $\sim 0.1~M_{\odot}$ looks 
as the most realistic estimate to date and suggests that MACHOs 
are brown dwarfs.

\section{Mass moment method}
 
The most appropriate way to compute the average mass and other
important properties of MACHOs is to use
the method of mass moments developed by De R\'ujula et al. \cite{kn:Derujula}.
The mass moments $<\mu^m>$ are 
related to $<\tau^n>=\sum_{events} \tau^n$,
with $\tau \equiv (v_H/r_E) T$, as constructed
from the observations 
($v_H = 210~{\rm km}~{\rm s}^{-1}$,
$r_E = 3.17 \times 10^9~{\rm km}$ and $T$ is the duration of an event
in days). We consider only 6
out of the 8 events observed by the MACHO group during their first two years.
In fact, the two disregarded events are 
a binary lensing and one which is rated as marginal.
The ensuing mean mass is
\footnote{When taking for the duration $T$ the values 
corrected for ``blending'', we get as average mass 0.34 $M_{\odot}$.
We thank D. Bennett for driving our attention on this point.}
$<\mu^1>/<\mu^0>=0.27~M_{\odot}$, assuming a standard spherical halo model.

Although this value 
is marginally consistent with the result of the MACHO team,
it definitely favours a lower average MACHO mass.
For the fraction of the local
dark mass density detected
in the form of MACHOs, we find $f \sim 0.54$, which compares quite well
with the corresponding value ($f \sim 0.45$) calculated
by the MACHO group in a different way. 

\section{Formation of dark clusters}

A major problem concerns the formation of MACHOs, as well
as the nature of the remaining amount of dark matter in the galactic halo.
We feel it hard to conceive a formation mechanism which transforms with 100\%
efficiency hydrogen and helium gas into MACHOs. Therefore, we expect
that also cold clouds (mainly of $H_2$) should be present in the 
galactic halo. Recently, we have proposed
a scenario \cite{de}
in which dark clusters of MACHOs and cold molecular 
coulds naturally form in the halo at galactocentric distances
larger than 10-20 kpc, with the relative abundance possibly
depending on the distance.

The evolution of the primordial proto globular cluster clouds
(which make up the proto-galaxy)
is expected to be very different 
in the inner and outer parts of the Galaxy, depending on the decreasing 
ultraviolet flux (UV) from the centre
as the galactocentric distance $R$ increases.
In fact, in the outer halo no substantial $H_2$ depletion should 
take place, owing to the distance suppression of the UV flux.
Therefore, the clouds cool and fragment - the process stops when the 
fragment mass becomes $\sim 10^{-2} - 10^{-1}~M_{\odot}$.
In this way dark
clusters should form, which contain brown dwarfs  
and also cold $H_2$ self-gravitating cloud, 
along with some residual diffuse gas (the 
amount of diffuse gas inside a dark cluster has to be low, for otherwise it 
would have been observed in the radio band).

We have also considered several observational tests for our model 
\cite{de,di}.
In particular, a signature for the presence of 
molecular clouds in the galactic halo should be a $\gamma$-ray flux 
produced in the scattering of high-energy cosmic-ray protons on $H_2$.
As a matter of fact, an essential
information is the knowledge of the cosmic ray flux in the halo. Unfortunately,
this quantity is unknown and the only available 
information comes from theoretical considerations.
Nevertheless, we can make an estimate of the expected $\gamma$-ray flux
and the best chance to detect it is provided
by observations at high galactic latitude.
Accordingly, we find a $\gamma$-ray flux (for $E_{\gamma}>100$ MeV)
$\Phi_{\gamma}(90^0) \simeq ~1.1 \times 10^{-6}~\epsilon f$ photons cm$^{-2}$
s$^{-1}$  sr$^{-1}$ ($\epsilon$ is a unknown parameter which takes into 
account the degree of confinement of the cosmic rays in the halo,
whereas $f$ stands for the fraction of halo dark matter in the form of
cold molecular gas).
This flux should be compared with the measured
value for the diffuse background of
$0.7-2.3\times 10^{-5}$ photons cm$^{-2}$ s$^{-1}$  sr$^{-1}$. Thus, 
there is at present no contradiction with observations.
Furthermore, an improvement of sensitivity for the next generation of 
$\gamma$-ray detectors will either discover the effect in 
question or yield more stringent limits on $\epsilon f$.\\


\begin{thebibliography}{99}
\bibitem{pa} B. Paczy\'nski, Astrophys. J. {\bf 304}, 1 (1986).
\bibitem{kn:Derujula1} A. De R\'ujula, Ph. Jetzer and E. Mass\'o,
Astron. Astrophys. {\bf 254}, 99 (1992). 
\bibitem{al} C. Alcock et al., Nature {\bf 365}, 621 (1993);
astro-ph 9606165.
\bibitem{au} E. Aubourg et al., Nature {\bf 365}, 623 (1993).
\bibitem{je} Ph. Jetzer, Helv. Phys. Acta {\bf 69}, 179 (1996).
\bibitem{kn:Ingrosso} F. De Paolis, G. Ingrosso and Ph. Jetzer,
Astrophys. J. {\bf 470}, 493 (1996).
\bibitem{kn:Derujula} A. De R\'ujula, Ph. Jetzer and E. Mass\'o,
Mont. Not. R. Astr. Soc. {\bf 250}, 348 (1991).
\bibitem{de} F. De Paolis, G. Ingrosso, Ph. Jetzer and M. Roncadelli,
Phys. Rev Lett. {\bf 74}, 14 (1995);
Astron. Astrophys. {\bf 295}, 567 (1995);
Comments on Astrophys. {\bf 18}, 87 (1995);
Astrophys. and Space Science {\bf 235}, 329 (1996);
Int. J. Mod. Phys. {\bf D5}, 151 (1996).
\bibitem{di} F. De Paolis, G. Ingrosso, Ph. Jetzer, A. Qadir and M. Roncadelli,
Astron. Astrophys. {\bf 299}, 647 (1995).


\end{thebibliography}
\end{document}